# Atomic layer engineering of perovskite oxides for chemically sharp heterointerfaces

Woo Seok Choi[1], Christopher M. Rouleau[1,2], Sung Seok A. Seo[1,3], Zhenlin Luo[4,5], Hua Zhou[6], Timothy T. Fister[4], Jeffrey A. Eastman[4], Paul H. Fuoss[4], Dillon D. Fong[4], Jonathan Z. Tischler[6], Gyula Eres[1], Matthew F. Chisholm[1], and Ho Nyung Lee[1]*

[1]*Materials Science and Technology Division, Oak Ridge National Laboratory, Oak Ridge, TN 37831, USA*
[2]*Center for Nanophase Materials Sciences, Oak Ridge National Laboratory, Oak Ridge, TN 37831, USA*
[3]*Department of Physics and Astronomy, University of Kentucky, Lexington, KY 40506, USA*
[4]*Materials Science Division, Argonne National Laboratory, Argonne, IL 60439, USA*
[5]*National Synchrotron Radiation Laboratory, University of Science and Technology of China, Hefei 230029, People's Republic of China*
[6]*X-ray Science Division, Advanced Photon Source, Argonne National Laboratory, Argonne, IL 60439, USA*

*E-mail: hnlee@ornl.gov

The two-dimensional (2D) conducting interface between $LaAlO_3$ (LAO) and $SrTiO_3$ (STO) – both perovskite oxide band insulators – has attracted immense attention since its discovery by Ohtomo and Hwang.[1] Several unexpected physical properties have been experimentally observed,[2,3] which stimulated numerous theoretical investigations.[4-6] However, despite intense research, the exact origin of such remarkable behavior remains to be understood. While the polar catastrophe model at the interface between a polar (LAO) and a non-polar (STO) oxide provides an appealing explanation for the characteristics of this system,[7] the role of extrinsic factors such as oxygen vacancies and/or cation intermixing in producing similar behavior cannot be ruled out.[8-15]

Pulsed laser deposition (PLD) is the most frequently used method for fabricating LAO/STO heterostructures. It is generally understood that, among the various growth parameters, the background gas pressure and the substrate temperature have the greatest influence on the film quality. Under optimal conditions – *i.e.*, when the oxygen partial pressure $P(O_2)$ is used to tune the kinetic energy of growth species in the laser-induced plume – single crystal thin films with atomically smooth surfaces can be grown.[16] However, at less than optimal growth conditions, the highly energetic plume-substrate interactions can produce undesirable departures from stoichiometry.[17,18] As summarized in Table 1 of Supporting Information, the PLD growth conditions used in published reports on the LAO/STO interface formation were not uniform, suggesting that the differences in the growth parameters might contribute to inconsistent observations and/or interpretations of interface conductivity. Most importantly, compared to conventional PLD growth conditions for typical perovskite oxides,[19,20] many LAO/STO heterostructures were grown at unusually low $P(O_2)$ and rather high temperatures.

The selective use of low $P(O_2)$ in the early stage of LAO/STO growth raises an intriguing question: what if instead very high $P(O_2)$ was used to grow the LAO film? It was shown that LAO/STO growth at low $P(O_2)$ easily generates a high concentration of oxygen vacancies,[8-11] and post-annealing in a high-pressure oxygen ambient was found to reduce their number.[3,8,9] Interestingly, some previous work reported good metallic conduction even after post-annealing, suggesting that the contribution of oxygen vacancies to the electronic conduction is negligible.[1,7,21] It should be further noted that the energetic process of PLD – *i.e.*, species are transported from the target to the substrate with a high kinetic energy – may substantially influence the formation of cation off-stoichiomety, resulting in a chemically broadened interface.[12-15] Here again, the $P(O_2)$ should have a strong influence on interface formation, but only a few studies explored high $P(O_2)$ growth of LAO/STO.[3,11,12] Furthermore, the subtle effects of $P(O_2)$ on the chemical composition of the interface and its structure, as well as its influence on subsequent growth have not been studied systematically.

Here, we show that $P(O_2)$ plays a dramatic role in determining the properties of LAO/STO at the initial stage of the growth. While low $P(O_2)$ growth resulted in a mostly relaxed LAO layer, inserting only a single monolayer (*i.e.*, one unit cell, u.c.) of LAO that was grown at high $P(O_2)$ was found to prevent strain relaxation. Analysis of these films using a combination of real time and post growth characterization methods, including x-ray diffraction (XRD) and electron microscopy/spectroscopy (see Supporting Information for experimental details) reveals that an interface engineered LAO film has a sharper interface with improved crystallinity when compared to LAO films grown without interface engineering. This result implies that $P(O_2)$ is instrumental for preventing cation intermixing at the interface under well-optimized laser conditions.

In heteroepitaxy, there exists a critical thickness above which the lattice of the thin film relaxes to relieve the epitaxial strain. In the case of LAO/STO, the LAO film is under tensile strain with a -2.9% mismatch. According to the mechanical equilibrium theory and empirical results, the critical thickness lies at around 2-10 nm.[15,22,23] Therefore, we have grown relatively thick films to explore whether interface mixing influences the strain relaxation. We specifically note that the growth of relatively thick LAO on STO has seldom been explored before,[14] even though it offers profound insight into the effect of strain on the interface conductivity.

Figure 1 shows XRD reciprocal space mapping (RSM) results of several 125 u.c. or ~50 nm thick LAO films on STO prepared at different $P(O_2)$. Note that this thickness is well above the critical thickness mentioned above. When a LAO/STO heterostructure was fabricated at $P(O_2) = 10^{-6}$ Torr, the resulting film revealed a significant strain relaxation, as confirmed by a RSM around the STO 114 Bragg reflection (Figure 1a). Since the film grown at such low $P(O_2)$ should have abundant oxygen vacancies, post-annealing in an oxygen atmosphere was used to compensate them. While *in-situ* post-annealing (immediately following the deposition at 700 °C in 50 mTorr of $O_2$ for 10 minutes) made the film highly insulating,[24] it had no effect on the relaxed crystal structure of LAO (Figure 1b). Next, we grew LAO films at high $P(O_2)$ ($10^{-2}$ Torr). Interestingly, the RSM in Figure 1c shows only a very weak LAO peak that is fully strained to the substrate. As illustrated schematically in Figure 1c, the weak peak intensity was caused by severe delamination and cracking of the film (see Supporting Information for a Nomarski image) as a consequence of elastic twin formation. The latter is a characteristic of LAO crystals. It is interesting to note that delamination has not been observed in films grown at low $P(O_2)$ (Figure 1a).

The behavior of the two films grown at these $P(O_2)$ extremes indicates that the initial growth of LAO on STO dictates the properties of the rest of the LAO film. Guided by this result, we grew a film using a two-stage process: first, one monolayer of LAO was grown at $P(O_2) = 10^{-2}$ Torr as a 'shielding layer,' and then the rest (*i.e.*, 124 u.c. in thickness) was grown at $P(O_2) = 10^{-6}$ Torr without interruption. The growth condition for the second stage was kept exactly the same as that for the film shown in Figure 1a. Surprisingly, the use of the single u.c. thick shielding layer resulted in a film coherently strained to the STO substrate (Figure 1d). In this regard, much thicker shielding layers were no more effective than a single monolayer one. For example, as shown in Figure 1e, a 20 u.c. thick shielding layer resulted in a similar outcome to that of films having only a one u.c. thick shielding layer. Therefore, it is reasonable that the lattice relaxation observed in the present case might be understood in terms of $P(O_2)$-dependent cation intermixing, as schematically illustrated in Figure 1f. If a shielding layer grown at high $P(O_2)$ can substantially reduce the intermixing, the resulting thin films may be able to sustain the tensile strain induced from the substrate (to at least 50 nm in the present case). Note that this thickness exceeds the theoretical value, as has also been observed in other oxide films.[19]

It should be noted that the oxygen vacancies generated in the STO substrate might also result in lattice expansion. However, since the oxygen vacancy compensated (Figure 1b) sample shows a similar relaxation behavior when compared to the low $P(O_2)$ grown LAO/STO, it is clear that the strain relaxation occurs during film growth, and post-annealing does not remove the dislocations formed to relieve the strain.

Specific results illustrating the effect of $P(O_2)$ on cation intermixing were obtained from nominally seven u.c. thick LAO films characterized by surface XRD rod scans at the Advanced Photon Source. The specular 00*L* rods were measured for the films grown at $P(O_2) = 10^{-6}$ Torr (Figure 2a and 2d), $P(O_2) = 10^{-6}$ Torr but with the first u.c. grown at $P(O_2) = 10^{-2}$ Torr (*i.e.*, a one u.c. thick shielding layer, Figure 2b and 2e), and $P(O_2) = 10^{-2}$ Torr (Figure 2c and 2f). Coherent Bragg Rod Analysis (COBRA)[13,25] was used to extract the one-dimensional electron density profiles, and the bootstrap method was used to determine the error bars.[26] The composition profiles are shown in Figure 2a-2c, where the electron density of the STO substrate is shown on the left side of the interface (located at layer number 0), and the electron density of the LAO film is shown on the right. As seen in Figure 2a, the number of electrons per layer within the substrate was larger near the interface for the film grown at low $P(O_2)$. This is indicative of La incorporation into the top of $SrTiO_3$. In contrast, the film grown at high $P(O_2)$ exhibited no such behavior (Figure 2c); the interface between the substrate and the film was abrupt, although the surface of the LAO was rougher. The effect of the one u.c. shielding layer is shown in Figure 2b. Here, La was not incorporated into STO, although some cation intermixing was observed in the first few u.c. of the film. Moreover, as shown in Figure 2d-2e, the change in the out-of-plane lattice spacing between the A-site cations across the interface supports our scenario as schematically represented in Figure 1f. The film grown at low $P(O_2)$ showed measurable lattice expansion in the STO substrate near the interface (Figure 2d). This expansion of the A-site lattice was mostly suppressed in the sample with a one u.c. shielding layer (Figure 2e), and negligible expansion was seen for the sample grown at $P(O_2) = 10^{-2}$ Torr (Figure 2f). Thus, there appears a correlation between intermixing in STO and a local expansion of the lattice.

The crystalline quality and the chemical abruptness of LAO/STO were further studied by Z-contrast scanning transmission electron microscopy (Z-STEM) equipped with electron energy loss spectroscopy (EELS). Figure 3a and 3b, respectively, show EELS elemental profiles of the integrated intensity of the La $M_{4,5}$ and Ti $L_{2,3}$ edges across the interface for LAO films (10 nm) shown in Figure 3c-3e. The film grown at high $P(O_2)$ clearly showed a smaller interface width, as determined from both the La and Ti profiles in Figure 3a and 3b. This interface profile was nearly identical to the simulated profile from the inelastic scattering model for a hypothetical, chemically abrupt LAO/STO interface.[27] The slight increase in interfacial width for the real system reflects the thermodynamic necessity for a diffuse interface between two coexisting phases and possible chemical effects driven by the interfacial screening charge that arises to compensate electric fields in polar materials.[28,29] The films grown with low $P(O_2)$ had broader interfaces defined mostly by the diffusion of Ti. Moreover, we found that the interface quality of the sample with a one u.c. shielding layer was also improved as compared to the sample entirely grown at $P(O_2) = 10^{-6}$ Torr. It should be noted that because of the very different character of the two measurement techniques (*e.g.* COBRA is a macroscopic tool, whereas Z-STEM-EELS is a microscopic one), there might remain some discrepancies in describing the detailed atomic profile. Nevertheless, the good qualitative agreement between COBRA and Z-STEM/EELS indicates that tuning $P(O_2)$ at the earliest stage of LAO/STO growth is critical in controlling the interface's structural quality.

To determine how the laser plume energetics affects interfacial mixing, we used real time ion probe measurements to characterize the kinetic energy distribution of ions in the plume at various $P(O_2)$.[30] Figure 4a shows the ion probe intensity as a function of time at different $P(O_2)$. At $P(O_2) = 10^{-6}$ Torr, the majority of ionic species arrived at the STO substrate within 2 $\mu$s after the laser pulse hits the target surface. As the $P(O_2)$ increased, the peak intensity of the signal decreased slowly. However, the arrival time of the ionic species on the substrate surface did not change substantially up to nearly $10^{-3}$ Torr. At $P(O_2) = \sim 10^{-2}$ Torr, however, propagation of the plume started to change significantly, and it slowed to 2.5 $\mu$s. Further increases in $P(O_2)$ shifted the peak to longer arrival time, and eventually its intensity decreased drastically, indicating that only a small fraction of the ionic species in the laser plume arrived at the substrate because of strong scattering by the oxygen background gas.[11,18] The kinetic energy of the atomic species approaches ~125 eV and remains unchanged up to $10^{-3}$ Torr, and then decreases significantly above that pressure (inset of Figure 4a). This clearly indicates that the plume-substrate interaction has the largest effect on the LAO film growth at low $P(O_2)$, and is a key factor responsible for cation intermixing as observed previously.[12-14]

The interface formation was also monitored in real-time using time-resolved surface x-ray diffraction (SXRD). The data are presented in the form of 2D plots of diffraction intensity as a function of in-plane momentum transfer, $q$ (Figure 4b and 4c). In addition to the reflection high-energy electron diffraction (RHEED)-like specular oscillations for $q = 0$, the SXRD maps also show weak diffuse scattering side lobes for $q \neq 0$, which oscillate out of phase with respect to the specular oscillations. The $\Delta q$ value of the diffuse scattering peaks corresponds to a correlation length $l = 2\pi/\Delta q$ that describes characteristic island spacing on the growing surface.[31] Interestingly, LAO films grown at $10^{-6}$ Torr (Figure 4b) showed no diffuse scattering peaks and a delayed appearance of the first specular oscillation peak, which we attribute to an irregular growth behavior such as substantial implantation of ablated species into the STO substrate. In contrast, a LAO film grown at $10^{-2}$ Torr (Figure 4c) showed pronounced periodic oscillations for both the specular and diffuse scattering peaks that are characteristic of layer-by-layer growth behavior, associated with sharp interface formation. These growth kinetics data clearly illustrate that the quality of oxide interfaces can be improved significantly by using optimal $P(O_2)$.

Figure 5 shows optical transmittance spectra for the samples presented in Figure 1. Note that optical spectroscopy can probe inherent charge carrier dynamics without extrinsic contributions from electronic boundary conditions of *dc* transport measurements.[32] The large absorption in the low photon energy region from LAO/STO grown at $P(O_2) = 10^{-6}$ Torr (Figure 5a) was due to metallic Drude absorption, which may be attributed to oxygen vacancies.[24] Also note that the dips at ~1.6 and ~2.4 eV in the transmittance spectra originate from the characteristic inter-band transitions in STO treated under reducing condition.[24] The Drude absorption was completely removed by annealing the sample in an oxygen atmosphere as shown in Figure 5b. Furthermore, no discernible Drude absorption was observed in a thinner LAO film (25 u.c. was grown to prevent delamination) grown at $10^{-2}$ Torr (Figure 5c). This observation suggests that the high $P(O_2)$ grown LAO is indeed insulating and does not exhibit any sign of metallicity readily found in low $P(O_2)$ grown LAO/STO heterostructures.

Shielding the substrate also influenced the generation of conducting carriers in the STO substrate. Figure 5d and 5e show optical transmittance spectra of LAO films grown at $10^{-6}$ Torr with one and 20 u.c. thick shielding layers, respectively. Even though both samples revealed similar spectral shapes as compared to the sample without a shielding layer (note that the ~1.6 and ~2.4 eV dips from oxygen vacancy were still visible), the number of conducting carriers generated in the 'shielded' sample were altered depending on the thickness of the shielding layer.

The temperature dependent resistance curves shown in the inset of Figure 5 were qualitatively consistent with the optical transmittance spectra. While the LAO film grown at $P(O_2) = 10^{-6}$ Torr and the low $P(O_2)$ grown sample with one u.c. thick shielding layer grown at high $P(O_2)$ exhibited a similar transport behavior, the LAO film with a 20 u.c. thick shielding layer grown at high $P(O_2)$ showed a discernibly larger (by a factor of five) sheet resistance. The larger electrical resistance originated mainly from the decrease in the temperature independent sheet carrier density from ~$10^{17}$ cm$^{-2}$ to ~$2 \times 10^{16}$ cm$^{-2}$, as the temperature dependent carrier mobility was similar for all three samples as confirmed by Hall measurements.[24] The transport behavior of our LAO/STO, thus, suggests that the polar catastrophe model does not simply apply, or cannot generate a sufficient number of carriers at least in our samples. In fact, a theoretical consideration of various electrical boundary conditions using the modern theory on polarization also found that the LAO/STO interface is intrinsically insulating,[33] consistent with our experimental observation. Based on our results, oxygen vacancies seem to play a fundamental role in the conduction mechanism of LAO/STO. It is also worth mentioning that in typical bulk polar materials, such as ferroelectrics, the surface and interface dipoles become compensated when interfacing with an insulating layer, e.g. by forming stripe domains,[34] inducing oxygen vacancies,[29] or adsorbing surface contaminants.[5,35]

In summary, subtle changes in the interface quality and the physical properties of LAO/STO heterostructures were explored by controlling the oxygen background pressure during the initial stage of PLD. When a LAO film was grown at very low oxygen partial pressure, we found the structure and chemical composition of the interface to be modified drastically, resulting in a greatly relaxed strain state. On the other hand, even a single u.c. thick LAO (~0.4 nm) grown at a high pressure acted as an effective shielding layer, thereby preventing strain relaxation in a subsequently grown LAO film at low pressure – at least up to a thickness of 50 nm, a value far beyond the theoretical critical thickness. We attribute this improvement to greatly reduced chemical intermixing in the STO substrate when high oxygen partial pressures (> 10 mTorr) were used for growing the LAO films. However, the formation of oxygen vacancies could not be completely avoided when a low oxygen partial pressure was used during the subsequent LAO deposition. Nevertheless, the ability to tune the structural quality of the interface offers an additional 'knob' for controlling the physical properties of this interface and heterostructures in general. Therefore, this work provides crucial information on subtle effects that must be carefully considered when designing artificial thin-film heterostructures with intriguing properties that are expected to arise from chemically well-defined oxide interfaces.

ACKNOWLEDGMENT. This work was supported by the U.S. Department of Energy, Basic Energy Sciences, Materials Sciences and Engineering Division. The optical characterization by a spectrophotometer was performed at the Center for Nanophase Materials Sciences, which is sponsored at Oak Ridge National Laboratory by the Scientific User Facilities Division, Office of Basic Energy Sciences, U.S. Department of Energy. The use of the Advanced Photon Source was supported by the U.S. Department of Energy, Basic Energy Sciences, under Contract No. DE-AC02-06CH11357.

FIGURE CAPTIONS.

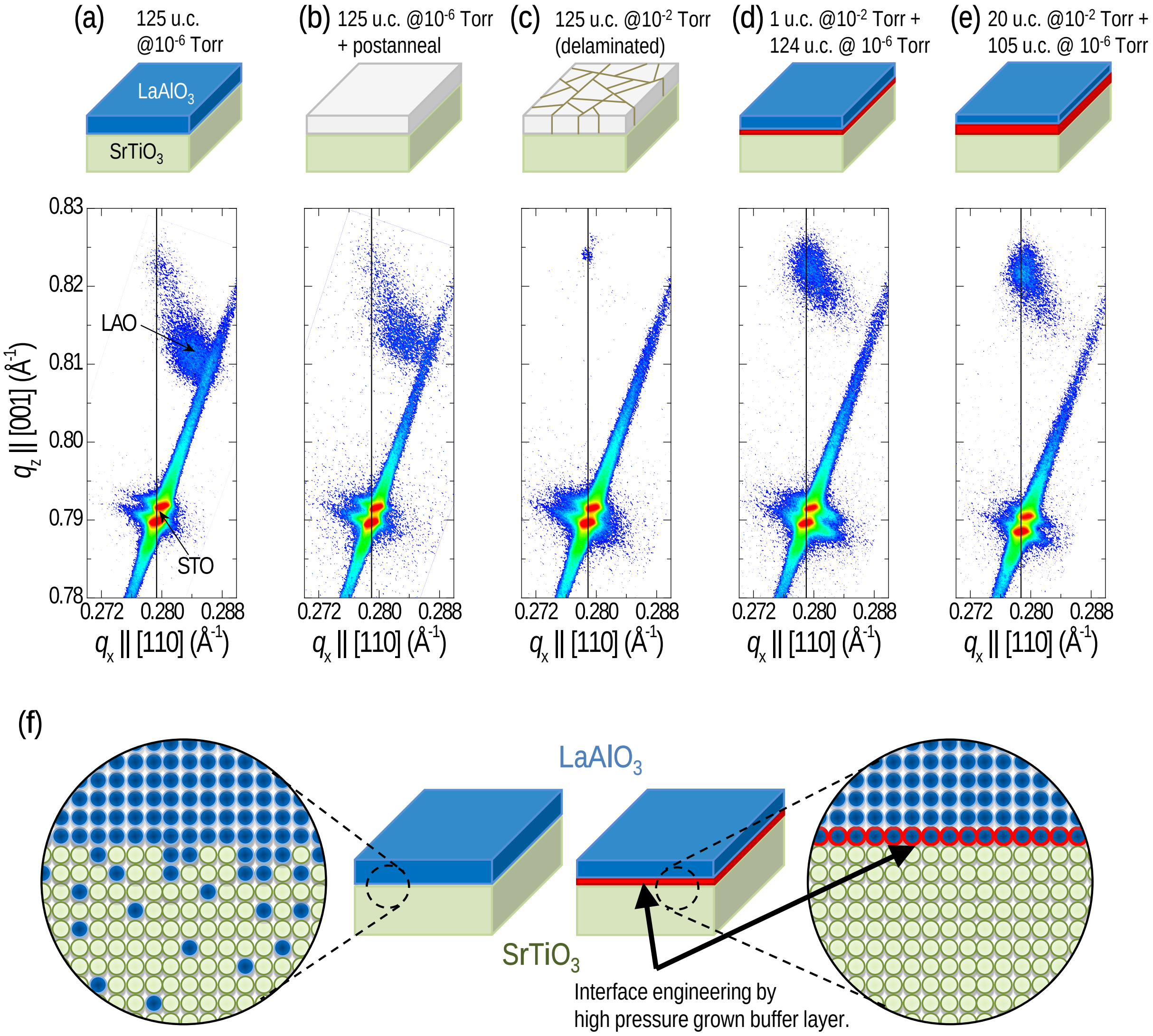

**Figure 1.** Strain and crystallinity of LAO thin films fabricated at different *P*($O_2$). (a-e) X-ray RSMs of LAO thin films around the substrate STO 114 Bragg peak. (Note that the reciprocal lattice unit is defined as $q = \lambda/2d$.) LAO films grown at *P*($O_2$) of (a) $10^{-6}$ Torr, (b) $10^{-6}$ Torr followed by an *in-situ* post-annealing at the growth temperature in 50 mTorr for 10 minutes, (c) $10^{-2}$ Torr, (d) $10^{-2}$ (one u.c.) + $10^{-6}$ Torr (124 u.c.), and (e) $10^{-2}$ (20 u.c.) + $10^{-6}$ Torr (105 u.c.). (f) Schematics of LAO films deposited on STO substrates. The left image represents a film

deposited at low $P(O_2)$ as shown in (a), where the highly-energetic plasma produces a chemically broadened interface. The right image illustrates the interface of the sample shown in (d). A single monolayer of LAO grown at high pressure serves as an efficient shielding layer, protecting the substrate surface from the energetic laser plume.

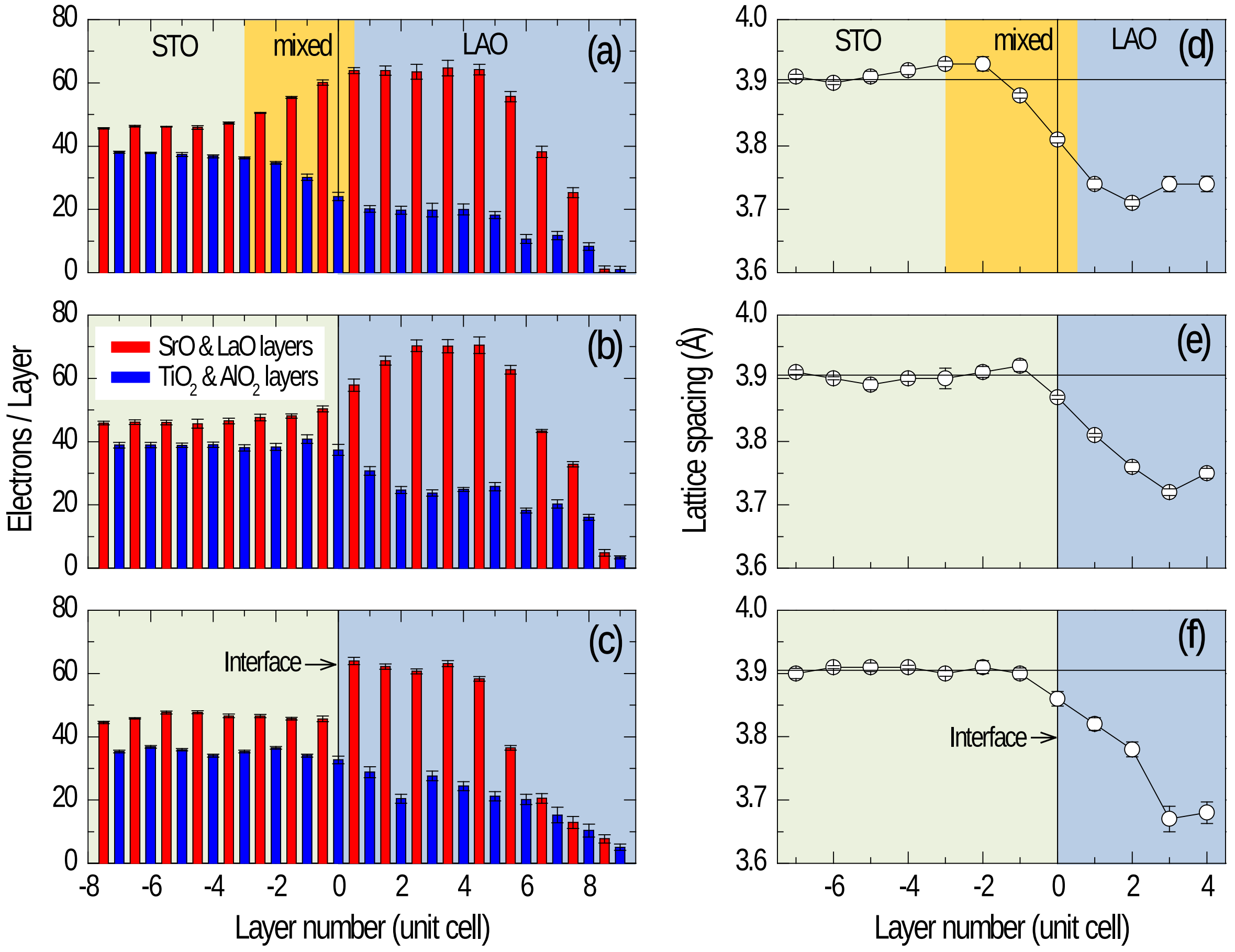

**Figure 2.** Room temperature electron density profiles for nominally seven u.c. thick LAO films grown at different $P(O_2)$ conditions. The $TiO_2$-terminated plane of the STO substrate is layer number 0. Electron density profile [(a), (b), and (c)] and the out-of-plane lattice spacing between the A-site cations measured across the interface [(d), (e), and (f)] for (a, d) a film grown at $P(O_2)$

= $10^{-6}$ Torr, (b, e) a film where the first u.c. was grown at $P(O_2)$ = $10^{-2}$ Torr and the remaining layers were grown at $P(O_2)$ = $10^{-6}$ Torr, and (c, f) a film entirely grown at $P(O_2)$ = $10^{-2}$ Torr, are respectively shown.

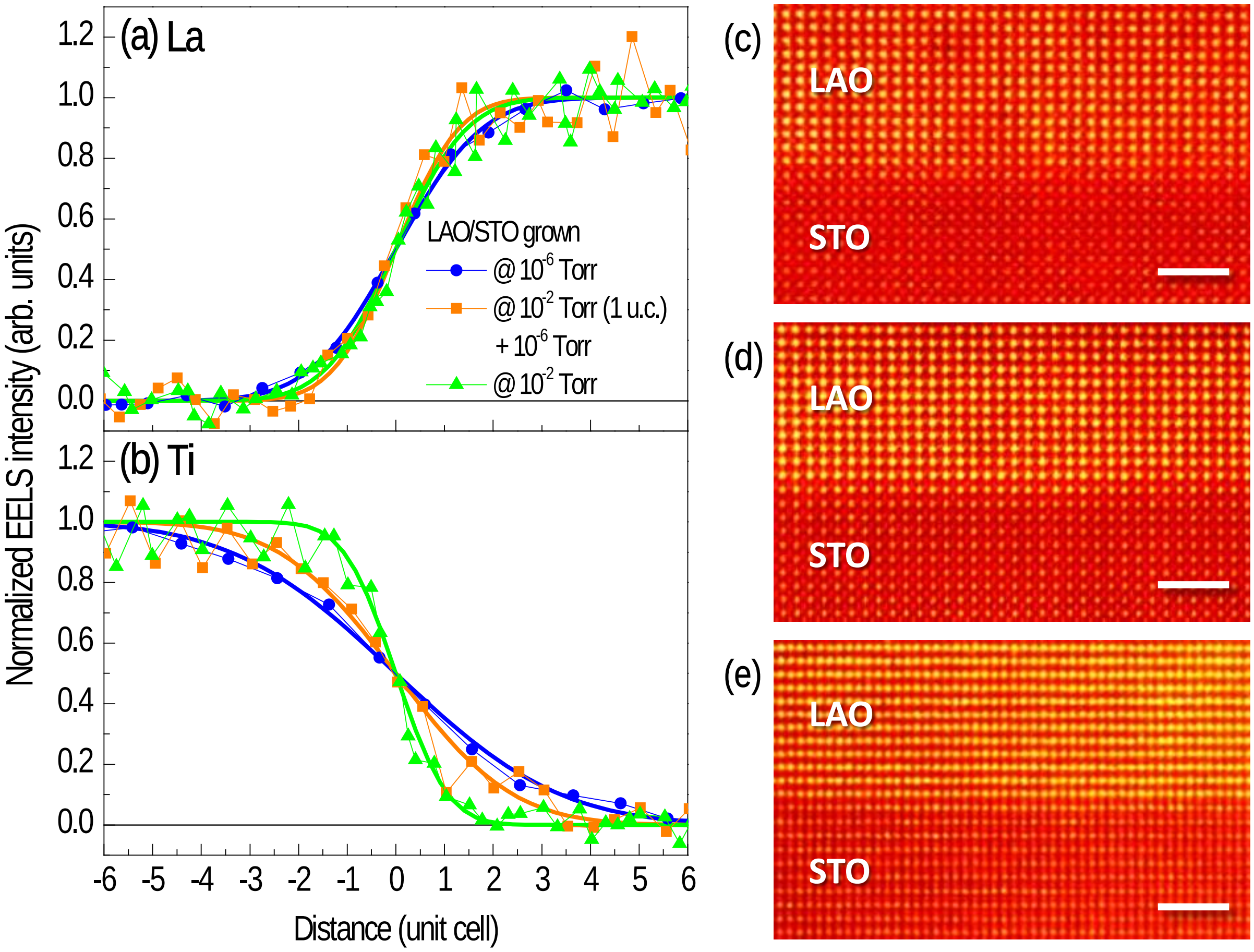

**Figure 3.** (a) La and (b) Ti EELS intensity profiles across the interface of LAO/STO grown at different $P(O_2)$. The symbols and the thick solid lines represent the experimental data and the error-function curve fit results for each profile, respectively. Z-contrast STEM images from LAO/STO grown at $P(O_2)$ of (c) $10^{-6}$ Torr, (d) $10^{-6}$ Torr with one u.c. shielding layer grown at $10^{-2}$ Torr, and (e) $10^{-2}$ Torr. Note that the images of (c) and (d) were taken along the [100] zone axis, while the one in (e) was along the [110] zone axis. The scale bars correspond to 2 nm.

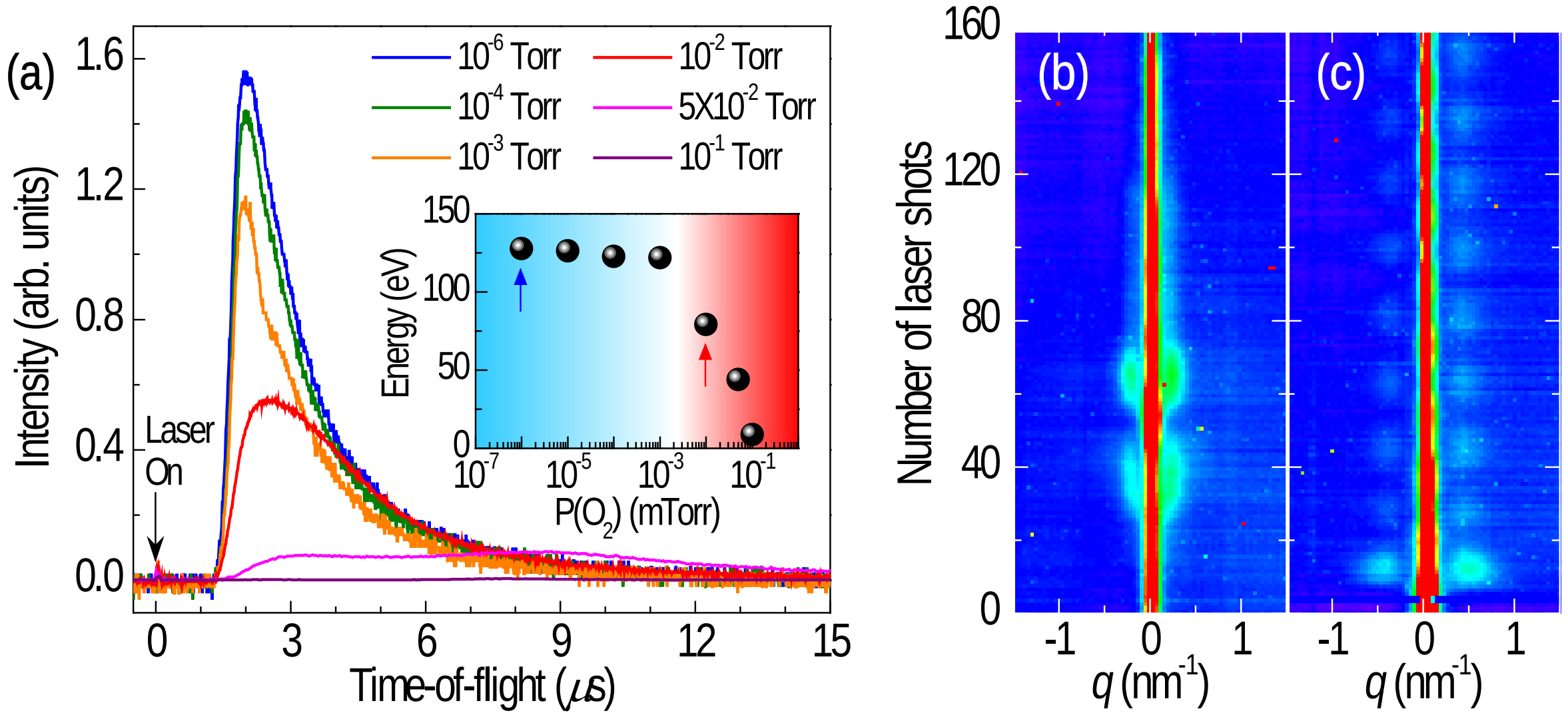

**Figure 4.** Real time kinetics during growth and structural characterization of LAO/STO. (a) Time-of-flight of ionic species in the plume measured by an ion probe at the substrate position as a function of $P(O_2)$. The inset shows the calculated kinetic energy of La arriving at the substrate surface at various $P(O_2)$. Real time SXRD recorded during PLD of LAO films grown at (b) $10^{-6}$ and (c) $10^{-2}$ Torr of $O_2$. The regularly spaced intensity oscillations of the diffuse scattering peak in (c) indicate a 2D layer-by-layer growth. On the other hand, as shown in (b), no pronounced 2D growth and a delayed appearance of the first oscillation peak were observed during the initial stage of the film growth at $P(O_2) = 10^{-6}$ Torr.

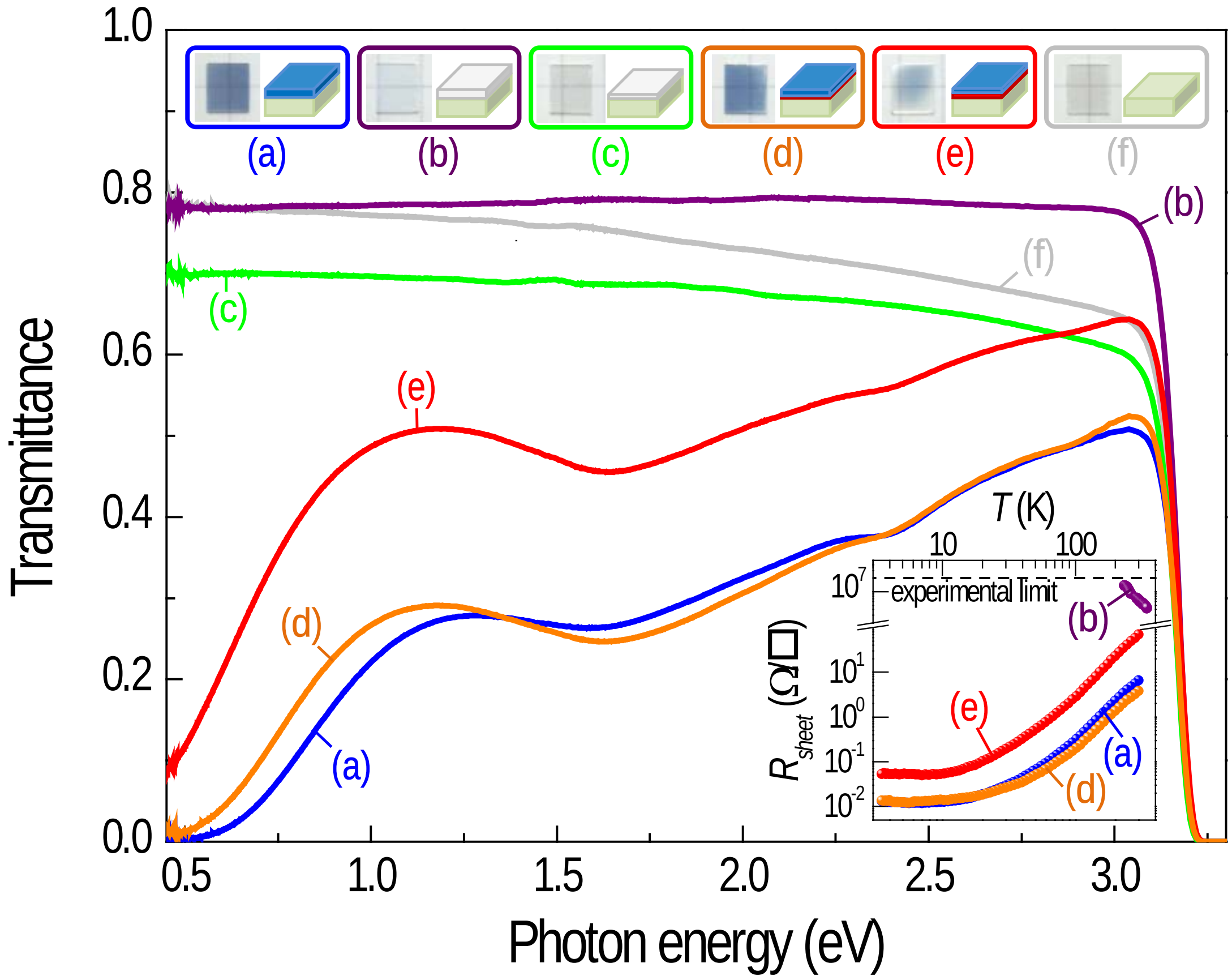

**Figure 5.** Optical transmittance spectra as a function of photon energy and corresponding photographs of LAO (125 u.c. in thickness)/STO heterostructures grown at $P(O_2)$ of (a) $10^{-6}$ Torr, (b) $10^{-6}$ Torr followed by an *in-situ* post-annealing at the growth temperature in 50 mTorr for 10 minutes, (c) $10^{-2}$ Torr (Note that this sample was only 10 nm-thick as a 50 nm-thick sample was delaminated), (d) $10^{-2}$ (one u.c.) + $10^{-6}$ Torr (124 u.c.), and (e) $10^{-2}$ (20 u.c.) + $10^{-6}$ Torr (105 u.c.). (f) A bare STO single crystal substrate is also shown for comparison. The inset shows temperature dependent sheet resistance curves for sample (a), (b), (d) and (e). Sample (c) had a similar insulating resistance as sample (b).

REFERENCES.

## Supporting Information

**Supporting Table 1.** Summary of early work on the LAO/STO growth by PLD.

| Reference | $T$[a)] [°C] | $P(O_2)$ [Torr] | $J$[b)] [J/cm$^2$] | $f$[c)] [Hz] | Post-annealing | $R_{xx}$[d)] @ low $T$ [Ω/□] |
|---|---|---|---|---|---|---|
| A. Ohtomo & H. Y. Hwang[S1] | 800 | $10^{-4}$ – $10^{-6}$ | 1 | 2 | - | ~$10^{-2}$ |
| M. Hujben *et al.*[S2] | 850 | $2\times10^{-5}$ | 1.3 | 1 | - | ~1 |
| S. Thiel *et al.*[S3] | 770, 815 | $1.5\times10^{-5}$ | - | - | 300 Torr, 1h @ 600 °C | ~1 |
| N. Nakagawa *et al.*[S4] | 750 | $10^{-5}$ | 3 | 5 | Oxygen, 4h @ 550 °C | - |
| W. Siemons *et al.*[S5] | 815 | $10^{-5}$ – $10^{-6}$ | 1.2 | 4 | @ various $T$ | ~$5\times10^{-2}$ - ~$10^3$ |
| A. Brinkman *et al.*[S6] | 850 | $1.9\times10^{-3}$ – $7.5\times10^{-7}$ | - | - | - | ~$10^{-2}$ - ~$10^5$ |
| P. R. Willmott *et al.*[S7] | 770 | $3.8\times10^{-6}$ | 1 | 10 | - | - |
| G. Herranz *et al.*[S8] | 750 | $7.5\times10^{-4}$ – $7.5\times10^{-7}$ | - | - | - | ~$10^{-2}$ - ~$10^8$ |
| A. Kalabukhov *et al.*[S9] | 800 | $7.5\times10^{-5}$ – $7.5\times10^{-7}$ | - | - | - | ~$10^{-2}$ - ~$10^3$ |
| N. Reyren *et al.*[S10] | 770 | $4.5\times10^{-5}$ | - | - | 300 Torr, 1h @ 600 °C | ~$10^2$ - $10^3$ |

[a)]Growth temperature; [b)]Laser fluence; [c)]Repitition rate; [d)]Sheet resistance.

## Thin Film Growth

PLD was used to grow LAO thin films on $TiO_2$-terminated single crystal (001) STO substrates. The samples were fabricated at 700 $^0$C in a range of $P(O_2)$ from $10^{-6}$ to $10^{-2}$ Torr. A KrF excimer laser ($\lambda$ = 248 nm) with a laser fluence of ~1 J/cm$^2$ and a spot size of 2 mm$^2$ was used for ablation of a single crystal LAO target. RHEED was used to monitor the surface structure and to control the film thickness with atomic-layer precision. According to our x-ray tube source diffraction data, all LAO films exhibited good crystallinity independent of the strain state as confirmed by rocking curve scans (with a full-width-at-half-maximum, $\Delta\omega \approx 0.04°$). Note that a typical $\Delta\omega$ is 0.03° from the 002 STO single crystal substrate peak, as measured with our curved mirror x-ray optics. Real time ion probe measurements were used to characterize the energetics and dynamics of the laser plume in various $P(O_2)$. The ion probe was located in the position of the STO substrate, and the TOF of the ionic species ejected from the target was measured.

## Sample Characterization

The LAO/STO heterostructures were characterized by several different techniques. Structural studies were carried out using an x-ray tube source (X'Pert, Panalytical Inc.) at Oak Ridge National Laboratory. COBRA was done with synchrotron radiation at the beamline 12-ID-D at the Advanced Photon Source (APS) at Argonne National Laboratory, using an x-ray energy of 24 keV at room temperature. We also conducted real time SXRD studies at the APS to obtain detailed information about the growth of the LAO layer during PLD. In addition to the specular intensity corresponding to RHEED-like growth intensity oscillations, the diffuse scattering component of the SXRD was also monitored to study the spatial distribution of islands occurring during layer filling. The effect of the $P(O_2)$ on the charge carrier generation was determined by measuring the optical transmittance spectra between 0.44 and 3.54 eV (CARY 5000, Varian Co.) at room temperature. A physical property measurement system (PPMS, Quantum Design Inc.) was used for *dc* transport characterization. Ohmic contacts of indium were ultrasonically soldered to the interface in Van der Pauw geometry and then wired using Au.

## Details on COBRA Analyses

The original data of Figure 3 is shown below (Figure S1). The excellent agreement between the converged COBRA results (red lines) and experimentally measured intensities (black circles) ensures the credibility of our data and analyses.

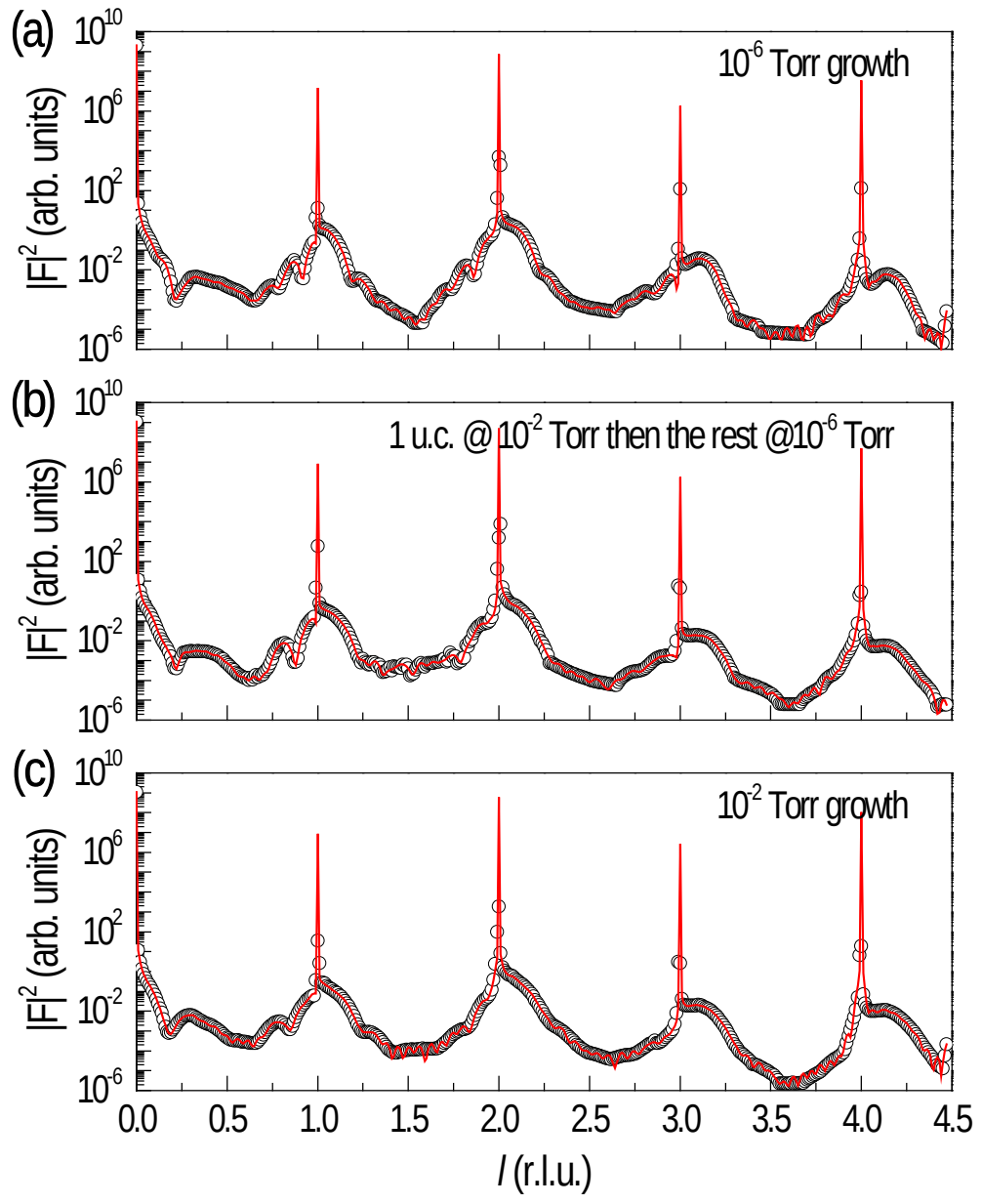

Figure S1. Scattered intensities along the specular 00*L* rod for the sample grown at (a) $10^{-6}$ Torr, (b) $10^{-2}$ Torr (one u.c.) then $10^{-6}$ Torr (the rest), and (c) $10^{-2}$ Torr of $O_2$.

The cumulative displacements of each atomic layer are shown in Figure S2. For the LAO film grown at $10^{-6}$ Torr of $O_2$ (Figure S2a), there was measurable lattice expansion in the STO side near the interface. A suppressed expansion was observed in the sample with one u.c. shielding layer (Figure S2(b)), and no expansion was observed for the sample grown at $10^{-2}$ Torr (Figure S2c). When these data were combined with those in Figure 3, there appeared a strong correlation between La incorporation into the STO and a local expansion of the lattice, especially for the sample grown at $P(O_2) = 10^{-6}$ Torr.

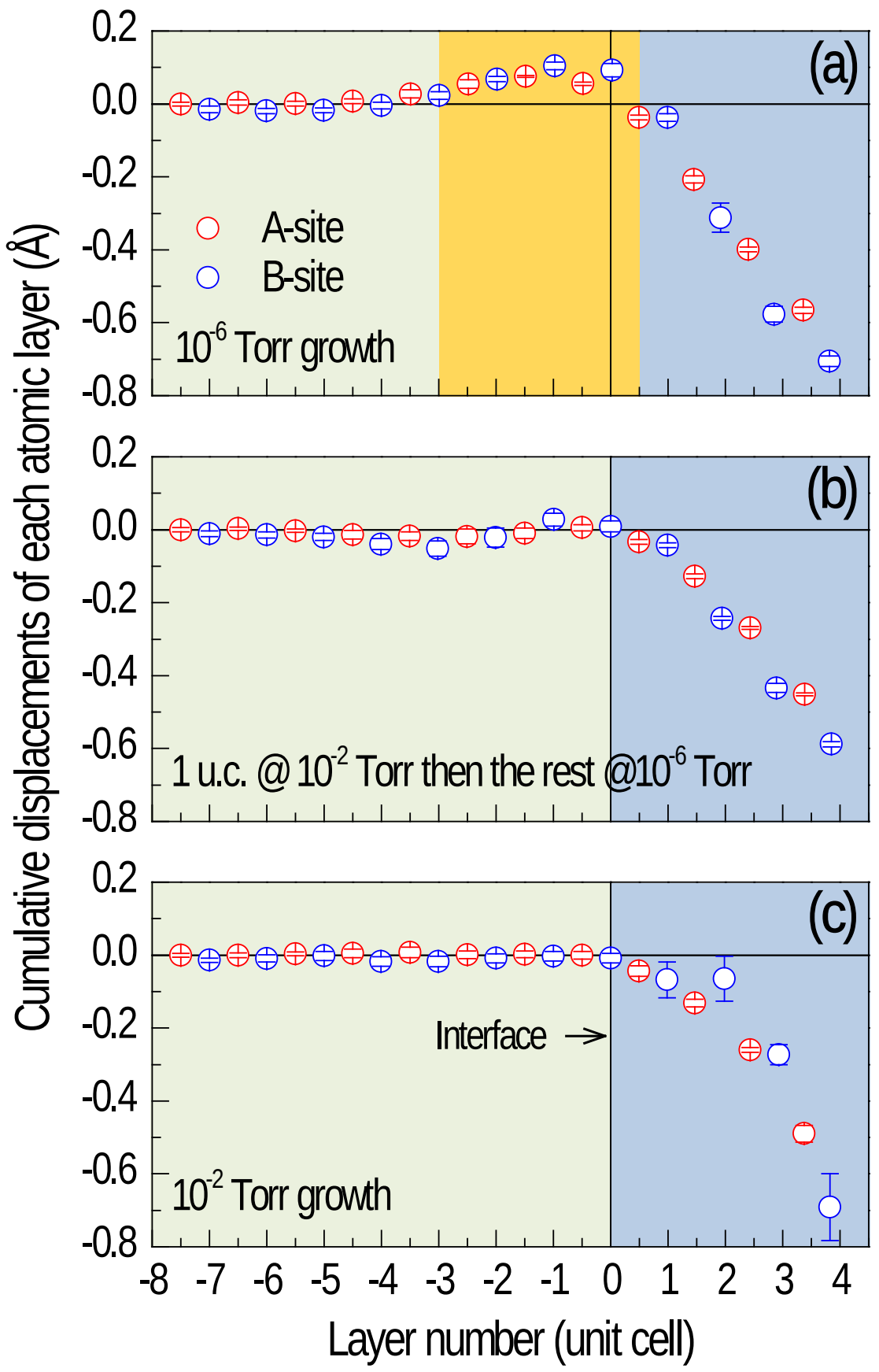

Figure S2. Cumulative displacements of the atoms measured with respect to the positions in the bulk STO lattice, with $c_{STO}$ = 3.905 Å for the LAO film grown at (a) $10^{-6}$ Torr, (b) $10^{-2}$ Torr (1 u.c.) then $10^{-6}$ Torr (the rest), and (c) $10^{-2}$ Torr of $O_2$. The vertical solid lines represent the nominal interface.

**Laser Condition (Spot Size and Fluence) Dependent Interface Quality**

Note that a recent report by Qiao *et al*. presents evidence of intermixing even in LAO grown at $P(O_2) = 10^{-2}$ Torr.[S11] However, in their work, an unusually high laser fluence (3 J/cm$^2$) and large laser spot (10 mm$^2$) were used. Since the kinetic energy of the laser plume is shown to be a critical factor for the intermixing in LAO/STO, our results are not inconsistent with that of Qiao *et al*.[11] This comparison further implies that the energetics of the PLD process that can modify the physical properties of a film are affected not only by the background pressure, but by other growth parameters as well. Thus optimal PLD growth conditions have to be determined through systematic tuning of the parameters.[S12,S13]

**Delamination of a Thick LAO Film Grown at High $P(O_2)$**

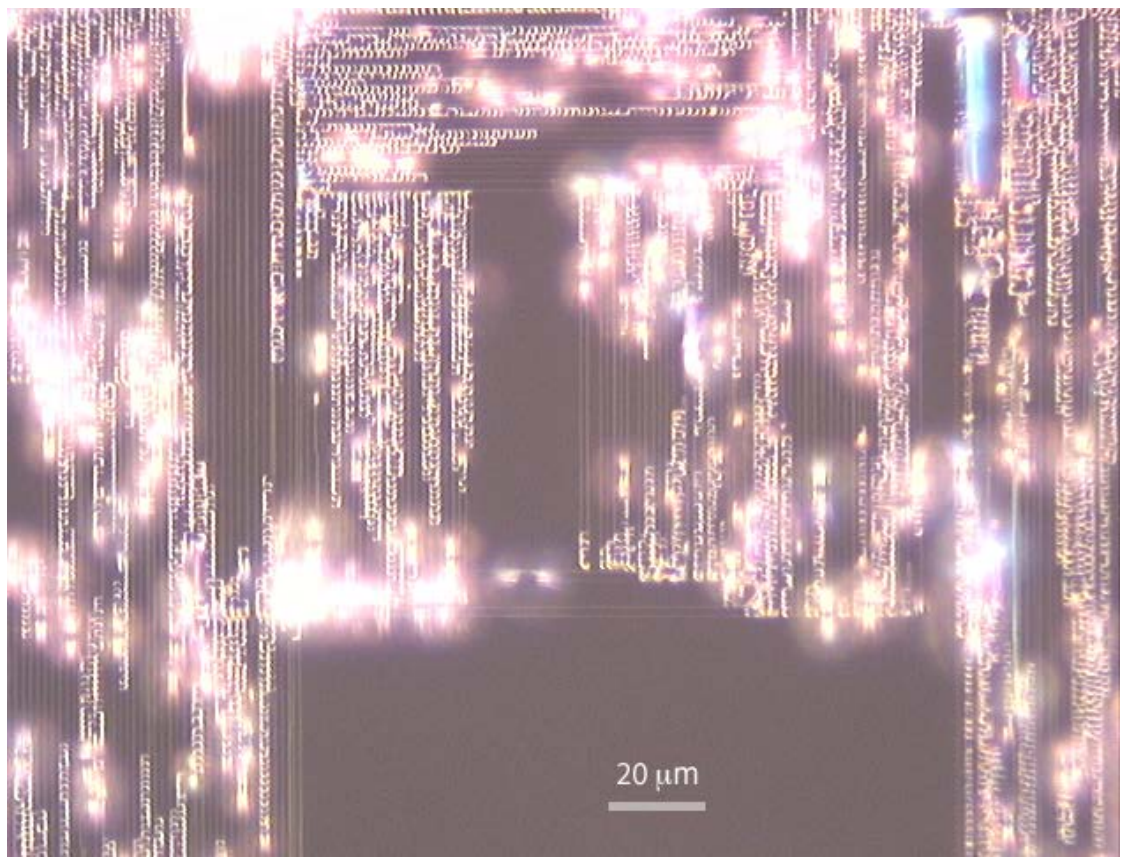

Figure S3. A Nomarski image of a delaminated LAO thin film (50 nm in thickness) on STO. The orthogonal arrays of delamination networks were formed to release the epitaxial strain when grown in high oxygen pressure (cf. Figure 1c).